\documentclass[twocolumn,twocolappendix]{aastex7}

\usepackage{amsmath}
\usepackage{subfigure}
\usepackage{CJKutf8}
\providecommand{\sorthelp}[1]{}

\definecolor{dgreen}{RGB}{26,148,49}

\shorttitle{Feature Intensity Mapping for PAHs}
\shortauthors{Cheng et al.}
\graphicspath{{./}{figures/}}

\begin{document}
\begin{CJK*}{UTF8}{bsmi}

\title{Feature Intensity Mapping: Polycyclic Aromatic Hydrocarbon Emission from All Galaxies Across Cosmic Time}

\author[0000-0002-5437-0504]{Yun-Ting Cheng (鄭昀庭)}
\affiliation{California Institute of Technology, 1200 East California Boulevard, Pasadena, CA 91125, USA}
\affiliation{Jet Propulsion Laboratory, California Institute of Technology, 4800 Oak Grove Drive, Pasadena, CA 91109, USA}
\email[show]{ycheng3@caltech.edu} 

\author[0000-0001-7449-4638]{Brandon S. Hensley}
\affiliation{Jet Propulsion Laboratory, California Institute of Technology, 4800 Oak Grove Drive, Pasadena, CA 91109, USA}
\email{brandon.s.hensley@jpl.nasa.gov}

\author[0000-0001-8490-6632]{Thomas S.-Y. Lai (賴劭愉)}
\affiliation{IPAC, California Institute of Technology, 1200 East California Boulevard, Pasadena, CA 91125}
\email{shaoyu@ipac.caltech.edu} 

\begin{abstract}
Line intensity mapping (LIM) is an emerging technique for probing the aggregate emission of a spectral line from all sources, without requiring individual detections. Through the wavelength--redshift relation, one can map the line-of-sight evolution of the line emission that traces the underlying large-scale structure in a spectral-imaging survey.  In this work, we present a new technique---\textit{feature intensity mapping}---as an extension of the LIM formalism to map broad spectral features in 3D, rather than the narrow emission lines typically targeted by LIM. By accounting for the convolution of spectral features with the instrument's spectral response across redshift, our technique enables simultaneous constraints on the redshift-dependent emission from multiple features. This approach enables 3D intensity mapping with some of the brightest features in the infrared spectra of galaxies: the polycyclic aromatic hydrocarbon (PAH) emission bands. We forecast the detectability of PAH signals using feature intensity mapping with the ongoing SPHEREx mission in the near-infrared and the proposed PRIMA mission in the far-infrared. We find that $S/N$ of $\gtrsim 10$ per redshift bin of widths $\Delta z = 0.1$ and $0.5$ can be achieved at $z < 0.5$ and $1 < z < 5$ with SPHEREx and PRIMA, respectively, for multiple PAH features, suggesting a promising prospect for mapping the aggregate PAH emission at cosmological distances with upcoming datasets.\footnote{\textcircled{c} 2025. All rights reserved.}
\end{abstract}


\section{Introduction}\label{S:introduction}
Intensity mapping (IM) is a technique to probe the cumulative emission from faint sources that lie below the individual detection threshold of imaging surveys \citep{Lagache:2005,Dole:2006,Cooray:2016}. By analyzing the aggregate signal, either in the monopole or in spatial fluctuations of the intensity field, IM enables statistical inference of the properties of unresolved source populations. Complementing conventional galaxy surveys, which only capture information from individually detected sources, IM offers a more comprehensive view of galaxy evolution. 

Line intensity mapping \citep[LIM; for reviews, see][]{Kovetz:2017,Bernal:2022} is a type of IM that targets specific spectral lines emitted by galaxies. Using narrow-band spectral-imaging mapping, LIM probes the integrated line emission from all sources across the sky. Through the redshift–frequency relation, the 3D (angular--spectral) LIM data cube directly traces the 3D spatial large-scale structure (LSS) of the Universe via cumulative line emission. LIM thus offers a powerful tool to study the 3D LSS, one of the most important cosmological observables, particularly at high redshift ($z \gtrsim 3$), where traditional galaxy surveys face limitations. Moreover, different spectral lines are sensitive to different physical processes within the interstellar medium (ISM) and intergalactic medium (IGM). Performing LIM across multiple lines thus offers a unique opportunity to study galaxy formation and evolution in various phases across cosmic time \citep{Sun:2019}. 

There are a number of ongoing and planned LIM experiments that target a wide range of the electromagnetic spectrum: the most extensively studied regimes are the radio for the 21 cm line \citep[for a review, see][]{Liu:2020} and the (sub)millimeter for $[$\ion{C}{2}$]$ and the CO rotational ladder \citep{Crites:2014,Keating:2015,Keating:2016,CONCERTOCollaboration:2020,Keating:2020,Ade:2020,Vieira:2020,Karkare:2022,Cleary:2022,CCAT-PrimeCollaboration:2023}. In the infrared (IR) regime, the recently launched NASA SPHEREx mission \citep{Crill:2020} is carrying out an all-sky near-infrared (NIR) spectral survey, making it well suited for LIM targeting multiple lines, including H$\alpha$, $[$\ion{O}{3}$]$, H$\beta$, and $[$\ion{O}{2}$]$ \citep{Cheng:2024}, among others \citep{Feder:2024}. For far-infrared (FIR) wavelengths, the proposed NASA Probe-class mission concept---the Probe Far-infrared Mission for Astrophysics \citep[PRIMA;][]{Glenn:2025}---is well suited for conducting such a large-area spectral-imaging survey. This would open new avenues for LIM using various FIR lines that trace star formation and black hole growth in galaxies (e.g., $[$\ion{N}{2}$]$, H$_2$, $[$\ion{S}{3}$]$, and $[$\ion{Si}{2}$]$; \citet{Moullet:2023}). 

In addition to spectral lines, among the most prominent features of the IR spectral energy distribution (SED) of galaxies are the emission bands from polycyclic aromatic hydrocarbons \citep[PAHs;][]{Leger:1984,Allamandola:1985,Smith:2004,Werner:2004}. Thes emissions arise from a series of vibrational modes, spanning wavelengths from 3~$\mu$m to 20~$\mu$m, with typical spectral widths ranging from $\sim$0.1 to a few microns \citep{Smith:2007}. PAH emission can contribute up to 20\% of the total IR luminosity and serves as an important tracer of dust content in galaxies \citep{Smith:2007,Lai:2020}, making PAH features well-suited targets for IM.

The PAH emission spectrum of a galaxy encodes a wealth of information about its ISM and the processes that shape it, including star formation history. The overall abundance of PAHs is now known to evolve systematically with metallicity \citep[e.g.,][]{Engelbracht:2005, Aniano:2020, Lai:2020, Whitcomb:2024}, hinting that gas-phase growth and destruction by ionizing photons may each play important roles in the PAH lifecycle \citep[see discussion in][]{Whitcomb:2024}. Indeed, PAHs are observed to be strongly depleted in \ion{H}{2} regions \citep{Chastenet:2019, Sutter:2024, Chastenet:2025}, with potential consequences for the PAH signal in compact, star-forming galaxies in the high-redshift Universe. The ratios of the feature strengths can indicate changes in size, charge state, and hydrogenation of PAHs, as well as the properties of the radiation field that illuminates them \citep[e.g.,][]{Tielens:2008, Maragkoudakis:2020, Draine:2021}. The PAH emission features have been used as tracers of star-formation \citep[e.g.,][]{ForsterSchreiber:2004, Peeters:2004, Shipley:2016, Lai:2020}, though it remains unclear with what fidelity PAH emission is correlated, or leads or lags, the evolution of the star-formation density with cosmic time. Certainly, correlations between PAH abundance and the specific star formation rate in galaxies have been observed \citep{Remy-Ruyer:2015, Nersesian:2019}.

Evidence has emerged that even the high-redshift Universe is enriched with PAHs. Emission in the 3.3\,$\mu$m PAH feature has been detected in a $z=2.0$ ultraluminous infrared galaxy (ULIRG) with Spitzer, with suggestions of the feature in three additional $z\simeq2$ ULIRGs \citep{Sajina:2009}. A second high redshift ($z=3.1$) detection of the 3.3\,$\mu$m feature was made in a lensed Lyman break galaxy with Spitzer \citep{Siana:2009}, and recently JWST has observed the 3.3\,$\mu$m PAH feature in a lensed $z=4.2$ galaxy \citep{Spilker:2023}. Moreover, the 2175\,\AA~ultraviolet extinction feature, often attributed to PAHs \citep[e.g.,][]{Joblin:1992, Duley:1998, Steglich:2010, Lin:2025}, has now been observed in multiple $z>6$ galaxies \citep{Markov:2023, Witstok:2023, Fisher:2025}. While JWST may be able to detect more bright PAH-emitting galaxies at cosmological redshifts, or detect high-redshift PAHs through rest-frame ultraviolet absorption in the 2175\,\AA\ feature, the vast majority of galaxies with PAHs will not be observed by JWST by virtue of being too faint, falling outside the small sky fraction JWST will ultimately observe at sufficient depth, and, at $z\gtrsim>2$, radiating most of their PAH emission at wavelengths longer than JWST can observe.

Applying IM techniques to the aggregate PAH emission background can complement observations on individual bright high-redshift galaxies. The aggregate background PAH emission has been detected through cross correlations. For example, \citet{Chiang:2019} detected cross-correlation signals up to $z\sim2$ between the Wide-field Infrared Survey Explorer (WISE) 12\,$\mu$m map and spectroscopic galaxies from the Sloan Digital Sky Survey, and attributed the signal to the cosmic PAH background. \citet{CordovaRosado:2024} detected a small angular scale cross correlation signal between broadband images from the WISE 12\,$\mu$m band and microwave temperature fluctuations measured by the Atacama Cosmology Telescope (ACT). They attributed the signal to the correlation between the cosmic infrared background in ACT and the PAH emission from unresolved galaxies in WISE. 

With narrow-band spectral-imaging surveys, the 3D distribution of PAHs across redshift can be extracted using a technique similar to LIM. However, emission lines are often modeled as delta functions in LIM due to the coarse spectral resolution compared to their intrinsic line widths. In contrast, the PAH features are relatively broad, such that emission from a given redshift is typically spread across multiple adjacent spectral channels. This complicates the conventional LIM framework in which each frequency channel corresponds to a distinct, non-overlapping redshift slice.

In this work, we introduce a new technique, which we refer to as ``feature intensity mapping,'' as an extension of LIM to broad spectral features such as PAHs in spectral-imaging surveys. By computing the convolution of the SED of a given feature with the observing filter profile as a function of redshift, our method models the full set of auto- and cross-power spectra between spectral channels. This framework captures both correlations from the same spectral feature across adjacent channels and those from different features tracing the same underlying LSS. We then use these measurements to infer the redshift-dependent total emission of all features in a Bayesian framework. This technique adopts a similar spirit to the \textit{Bayesian Multi-line Intensity Mapping} framework developed in \citet{Cheng:2024}.

The \textit{feature intensity mapping} technique enables 3D IM of one of the brightest IR SED features from PAHs, which would otherwise be inaccessible using the conventional LIM formalism. Furthermore, since PAHs are broad spectral features, the required spectral resolution is lower compared to LIM. This makes fast, wide-area mapping feasible even with low-resolution ($R \sim 10$) spectral imaging. This technique is developed in time for the upcoming SPHEREx observations in the $0.75$--$5\,\mu$m range, which will be sensitive to the 3.3\,$\mu$m PAH feature at $z \lesssim 0.5$ \citep{Zhang:2025}. The proposed PRIMA mission, targeting the $24$--$235$\,$\mu$m wavelength range, is also well suited for mapping various PAH features across cosmic time \citep{Bisigello:2024, Faisst:2025}. In this study, we apply our technique to forecast the sensitivity to the aggregate intensity of various PAH features for both SPHEREx and PRIMA. We find that high-significance detections, at the level of several tens of $\sigma$, for multiple PAH features can be achieved at $z\lesssim0.5$ for SPHEREx and $z\lesssim5$ for PRIMA. These forecasting results suggest the great promise of our feature intensity mapping technique for probing the 3D cosmic PAH background.

This paper is organized as follows. In Section~\ref{S:feature_intensity_mapping}, we introduce the formalism of the feature intensity mapping technique. Section~\ref{S:survey_setup} describes the assumed survey configurations for our SPHEREx and PRIMA forecast. The modeling of various spectral features is detailed in Section~\ref{S:signal_modeling}. We present our forecast results in Section~\ref{S:results} and provide further discussion in Section~\ref{S:discussion}. We summarize our conclusions in Section~\ref{S:conclusion}. Throughout this work, we assume a flat $\Lambda$CDM cosmology consistent with the measurements from Planck \citep{planck2016-l06}.

\section{Feature Intensity Mapping}\label{S:feature_intensity_mapping}
This section describes the formalism for \textit{feature intensity mapping}. We begin by introducing the intensity field arising from the aggregate galactic emission across cosmic time (Section~\ref{S:intensity_field}). The SED of each galaxy contains multiple spectral features, including emission lines and broad features such as the PAH emission features (see Section~\ref{S:signal_modeling} for details). We then present the formalism for computing the large-scale power spectrum of this emission field (Section~\ref{S:power_spectrum}). Our formalism largely follows the \textit{Bayesian Multi-line Intensity Mapping} framework presented in \cite{Cheng:2023}. With the formalism established, we introduce a (nearly) assumption-free parametrization to describe the redshift evolution of the average emission for each feature (Section~\ref{S:parametrization}) and outline the inference framework based on the Fisher information formalism (Section~\ref{S:inference}).

\subsection{Intensity Field}\label{S:intensity_field}
Consider a spectral feature $i$ with an intrinsic normalized spectral profile $S_{\rm int}(\lambda_{\rm rf})$ defined as
\begin{equation}
S_i(\lambda_{\rm rf}) = \frac{\lambda_{\rm rf} L_{\lambda}(\lambda_{\rm rf})}{L},
\end{equation}
where $\lambda_{\rm rf}$ denotes the rest-frame wavelength,\footnote{Throughout this manuscript, $\lambda$ and $\nu$ refer to the observed wavelength and frequency, while the rest-frame wavelength and frequency are denoted by $\lambda_{\rm rf}$ and $\nu_{\rm rf}$, respectively.} and the normalization factor $L$ is defined by $L \equiv \int d\lambda_{\rm rf}~L_\lambda(\lambda_{\rm rf})$, such that 
$S_i(\lambda_{\rm rf})$ is normalized: $\int d(\ln \lambda_{\rm rf})~S_i(\lambda_{\rm rf}) = 1$.

When observing this spectrum at redshift $z$ with a filter $\alpha$ characterized by a transmission curve $R_\alpha(\lambda)$, the filter-convolved normalized spectral profile $S_{\alpha,i}$ is given by\footnote{Note that this expression requires the filter transmission $R_\alpha(\lambda)$ to be defined as the photon response function.}
\begin{equation}
\begin{split}
S_{\alpha,i}(z) &= \frac{\int d\lambda~R_\alpha(\lambda)~\frac{\lambda L_{\lambda}(\lambda)}{L}}{\int d\lambda~R_\alpha(\lambda)} \\
&= \frac{\int d\lambda_{\rm rf}~R_\alpha((1+z)\lambda_{\rm rf})~S_i(\lambda_{\rm rf})}{\int d\lambda_{\rm rf}~R_\alpha((1+z)\lambda_{\rm rf})}.
\end{split}
\end{equation}

The total emission intensity field from spectral feature $i$ observed in filter $\alpha$ at an angular position $\hat{n}$ is given by
\begin{equation}
\begin{split}
\lambda I_{\lambda,i}(\alpha, \hat{n}) &= \int d\chi \int dL~\Phi_i(L, z, \hat{n}) D_A^2(z) \frac{L~S_{\alpha,i}(z)}{4\pi D_L^2(z)} \\
&= \int d\chi~S_{\alpha,i}(z)~M_{0,i}(z,\hat{n})~A(z),
\end{split}
\end{equation}
where $\chi$ is the comoving distance,\footnote{The redshift $z$ and comoving distance $\chi$ are used interchangeably throughout this manuscript.} $\Phi_i(L, z, \hat{n}) = dn/dL$ is the luminosity function of feature $i$ at redshift $z$ and angular position $\hat{n}$, and $D_A$ and $D_L$ are the comoving angular diameter distance and luminosity distance, respectively. We define the following quantities to simplify the expression:
\begin{equation}
\begin{split}
M_{0,i}(z,\hat{n}) &= \int dL~L~\Phi_i(L, z, \hat{n}), \\
A(z) &= \frac{D_A^2(z)}{4\pi D_L^2(z)}.
\end{split}
\end{equation}
Note that $M_{0,i}(z,\hat{n}) = dL_i/dV(z,\hat{n})$ represents the comoving luminosity density. We use $M_{0,i}(z)$ to denote the mean luminosity density averaged over angular positions.

The total intensity field from all spectral features is given by the sum of the emissions from each feature:
\begin{equation}
\lambda I_{\lambda}(\alpha, \hat{n}) = \sum_i \lambda I_{\lambda,i}(\alpha, \hat{n}).
\end{equation}

\subsection{Power Spectrum}\label{S:power_spectrum}
We measure the information from spectral-intensity maps in the spherical harmonics space, where the auto- and cross-angular power spectra $C_{\ell,\lambda\lambda'}$ characterize the covariance of the field. We focus on large-scale covariance on clustering scales, where the emission sources trace the underlying LSS, and we neglect higher-order effects such as redshift-space distortions and lensing magnification. We also ignore fluctuations from nonlinear clustering and Poisson noise. Since large-scale fluctuations can be fully described by a Gaussian probability distribution, the $C_{\ell,\lambda\lambda'}$'s capture all two-point information on large scales. This framework follows previous work \citep{Cheng:2023, Cheng:2024}.

The observed power spectrum is the sum of contributions from the signal (only the clustering component is considered) and noise:
\begin{equation}
\mathbf{C}_{\ell} = \mathbf{C}_{\ell}^{\rm clus} + \mathbf{C}_{\ell}^{\rm n}.
\end{equation}
Here, the boldface $\mathbf{C}_{\ell}$ denotes the angular power spectrum matrix of size $N_{\rm filt}\times N_{\rm filt}$, where $N_{\rm filt}$ is the number of observing filters, and each element is given by the cross spectra of maps observed in filters $\alpha$ and $\beta$, $\mathbf{C}_{\ell,\alpha\beta}$.

The clustering angular cross-power spectrum of two maps observed through filters $\alpha$ and $\beta$ is given by \citep{Cheng:2023}
\begin{equation}\label{E:Cl_clus}
\begin{split}
C_{\ell,\alpha\beta}^{\rm clus} =& \sum_{i,j}\int \frac{dk}{k}\frac{2}{\pi}k^3P(k) \\ 
&\cdot\int d\chi~S_{\alpha,i}(z)M_i(z)A(z)D(z)j_\ell(k\chi)\\
&\cdot\int d\chi'~S_{\beta,j}(z')M_j(z')A(z')D(z')j_\ell(k\chi'),
\end{split}
\end{equation}
where $P(k)$ is the present-day matter power spectrum, $D(z)$ is the linear growth factor, and $j_\ell$ is the spherical Bessel function. The quantity $M_i \equiv b_i(z)M_{0,i}(z)$ is the bias-weighted luminosity density, where $b_i(z)$ is the large-scale bias factor. In the large-scale linear regime, the power spectrum transfer function reduces to the scale-independent growth factor $D(z)$, the bias factor $b_i(z)$ is also scale-independent, and we use the linear matter power spectrum for $P(k)$.

Assuming white noise in each observing band with no cross-channel correlations, the noise power spectrum is given by
\begin{equation}\label{E:Cl_n}
C_{\ell,\alpha\beta}^{\rm n} = \sigma_{n,\alpha}^2~\Omega_{\rm pix}\delta^K_{\alpha\beta},
\end{equation}
where $\sigma_{n,\alpha}$ is the rms noise fluctuation in filter $\alpha$ per pixel, $\Omega_{\rm pix}$ is the pixel solid angle, and $\delta^K$ is the Kronecker delta.

\subsection{Parametrization}\label{S:parametrization}
Our main objective is to constrain the redshift evolution of the bias-weighted luminosity density for each feature: $M_i(z) = b_i(z)\,dL_i(z)/dV$. We aim to extract this information from the correlations in the spectral-intensity maps.

Following \cite{Cheng:2024}, we adopt a nearly assumption-free parametrization for $M_i(z)$, in which we characterize $M_i(z)$ at a set of redshift anchor points and linearly interpolate between them. This approach avoids imposing a specific analytical form with a limited number of free parameters, allowing for maximal flexibility in the redshift dependence of $M_i(z)$. Since $M_i(z)$ is expected to evolve smoothly with redshift, a small number of redshift anchors per feature is typically sufficient to describe its redshift evolution.

We select different sets of redshift anchors for different spectral features and observational configurations (see Section~\ref{S:feature_models} and Table~\ref{T:features} for the choices of anchors of each feature). The cosmological model is held fixed; therefore, the total number of free parameters corresponds to the sum of the number of redshift anchors across all features under consideration. In practice, some spectral features are likely to be highly correlated. Accounting for such correlations can effectively reduce the number of independent parameters and lead to improved constraints on these correlated features. We defer a detailed exploration of this aspect to future work.

\subsection{Inference}\label{S:inference}
We use the Fisher formalism to forecast parameter constraints. Given a set of spectral maps observed in multiple filters, and assuming a Gaussian likelihood, the Fisher matrix element for parameters $\theta$ and $\phi$ is given by
\begin{equation}\label{E:fisher}
\mathbf{F}_{\theta\phi} = \frac{1}{2}\sum_\ell n_\ell~ {\rm Tr}\left(\mathbf{C}_\ell^{-1}\frac{\partial \mathbf{C}_\ell}{\partial\theta}\mathbf{C}_\ell^{-1}\frac{\partial \mathbf{C}_\ell}{\partial\phi}\right),
\end{equation}
where the parameters correspond to the values of $M_i(z)$ at the redshift anchor points defined in Section~\ref{S:parametrization}. The factor $n_\ell$ denotes the number of available multipole modes in a given $\ell$ bin defined over the range $\ell \in [\ell_{\rm min}, \ell_{\rm max})$:
\begin{equation}\label{E:nell}
n_\ell = f_{\rm sky}\left(\ell_{\rm max}^2 - \ell_{\rm min}^2\right),
\end{equation}
with $f_{\rm sky}$ representing the survey sky fraction. We compute the parameter derivatives $\partial \mathbf{C}_\ell/\partial\theta$ using the automatic differentiation capabilities of the Python package \texttt{JAX} \citep{jax2018github}.

The inverse of the Fisher matrix yields the covariance matrix of the parameters:
\begin{equation}
\mathbf{C}_{\theta\phi} = \left(\mathbf{F}^{-1}\right)_{\theta\phi}.
\end{equation}

In Section~\ref{S:results}, we will also quantify the total signal-to-noise ratio ($S/N$) for detecting a spectral feature across all redshifts. The total $S/N$ for feature $i$ is defined as
\begin{equation}\label{E:SNR_tot}
\left(\frac{S}{N}\right)_{\rm tot,i} = \sqrt{\mathbf{m}_i^T\mathbf{C}_i^{-1}\mathbf{m}_i},
\end{equation}
where $\mathbf{m}_i$ is the vector of $M_i(z)$ values at the redshift anchor points for feature $i$, and $\mathbf{C}_i$ is the corresponding covariance matrix. This expression accounts for the covariance across redshift bins when evaluating the total sensitivity, while neglecting correlations between different spectral features.

One of the main challenges in LIM is interloper confusion, which occurs when multiple spectral lines from different line-of-sight distances are redshifted to the same observed wavelength, causing their signals to overlap in the LIM dataset. Our technique, extending the formalism of \citet{Cheng:2024}, addresses this issue by exploiting the distinct correlation structures of different lines in spectral-spatial space and by jointly fitting for multiple lines (or features, in our case). \citet{Cheng:2024} demonstrated that this framework enables an unbiased multi-line intensity mapping inference without the interloper issue. The approach adopted here inherits this robustness and is similarly capable of disentangling and mapping multiple broad spectral features from the data.

\section{Survey Setup}\label{S:survey_setup}
We demonstrate our feature intensity mapping technique by forecasting signal constraints with SPHEREx (Section~\ref{S:survey_setup_spherex}) and PRIMA (Section~\ref{S:survey_setup_prima}). Both missions are designed to perform large-scale spectral mapping in the infrared, where one or more PAH features fall within their observing windows, making them well suited for applying our method to probe a variety of spectral features across redshifts.

\subsection{SPHEREx}\label{S:survey_setup_spherex}

\begin{figure}[ht!]
\begin{center}
\includegraphics[width=\linewidth]{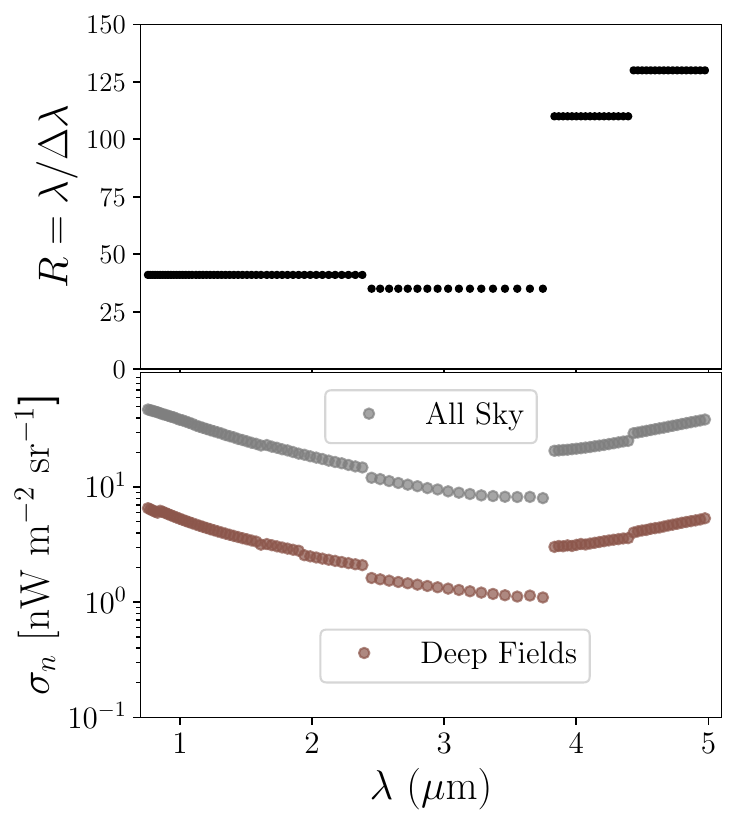}
\caption{\label{F:spherex_R_sign}Top: SPHEREx spectral resolution of each channel. Bottom: SPHEREx surface brightness sensitivity per spectral channel in a $6''.2$ sky pixel. The brown and gray points represent the SPHEREx all-sky and deep-field sensitivity, respectively.}
\end{center}
\end{figure}

The SpectroPhotometer for the History of the Universe, Epoch of Reionization, and Ices Explorer \citep[SPHEREx;][]{Crill:2020} is a NASA Medium Class Explorer mission that was successfully launched on March 11, 2025. Following the In-Orbit Checkout phase, SPHEREx began its two-year nominal science survey on May 1, 2025.

SPHEREx is conducting the first all-sky NIR spectro-imaging survey with a pixel size of $6''.2$, completing four full-sky passes over the nominal two-year mission duration. SPHEREx uses six broadband detector arrays spanning wavelengths from 0.75 to 5~$\mu$m, with low-resolution spectroscopy enabled by linear variable filters \citep{Korngut:2018}. The spectral coverage is divided into 17 channels per band, with varying spectral resolution: $R=41$ for bands 1--3 (covering 0.75--2.42~$\mu$m, totaling 51 channels), $R=35$ for band 4 (2.42--3.82~$\mu$m, 17 channels), $R=110$ for band 5 (3.82--4.42~$\mu$m, 17 channels), and $R=130$ for band 6 (4.42--5.00~$\mu$m, 17 channels).

Operating in a sun-synchronous orbit, SPHEREx visits the north and south ecliptic poles on every orbit, yielding integration time approximately 50 times deeper in those regions than the all-sky area \citep{Bryan:2025}. Consequently, SPHEREx will produce two deep-field mosaic maps of approximately 100~deg$^2$ each, optimized for studying the extragalactic background light.

In this work, we consider spectral-imaging maps from SPHEREx in both its all-sky and deep-field configurations. For the all-sky case, we adopt an observed sky fraction of $f_{\rm sky} = 70\%$ to account for Galactic masking. For the deep fields, we consider a total survey area of 200~deg$^2$. We use a Gaussian filter transmission profile with the FWHM given by the nominal spectral resolution. Figure~\ref{F:spherex_R_sign} presents the spectral resolution and the rms noise levels for both the all-sky and deep-field modes across the 102 SPHEREx channels. The noise levels are derived from the SPHEREx Sky Simulation \citep{Crill:2025}, which takes into account both instrumental noise and
the photon noise from various sources of emission, with the zodiacal light being the dominant source of the photon noise.

We adopt a multipole range of $50 < \ell < 350$ with 10 logarithmically spaced $\ell$ bins for our SPHEREx analysis, following the same setup as \citet{Cheng:2024}.
The minimum multipole ($\ell_{\rm min} = 50$) corresponds to the angular scale of SPHEREx’s $3.5^\circ$ field of view per detector. Fluctuations on larger angular scales will be partially suppressed due to the filtering of zodiacal light. The maximum multipole ($\ell_{\rm max} = 350$) is chosen to limit our analysis to large-scale modes where the power spectrum remains in the linear regime.

\subsection{PRIMA}\label{S:survey_setup_prima}

The PRobe far-Infrared Mission for Astrophysics \citep[PRIMA;\footnote{\url{https://prima.ipac.caltech.edu/}}][]{Glenn:2025} is a 1.8~m cryogenically cooled far-infrared space observatory, currently a proposed concept for the first NASA Probe-Class mission.

PRIMA is designed to include two instruments: an imaging instrument, PRIMAger \citep{Ciesla:2025}, and a spectrometer, FIRESS \citep{Bradford:2025}. PRIMAger consists of two hyperspectral imagers with low-resolution spectroscopy capability ($R=10$) enabled by linear variable filters, covering the wavelength range from 24 to 84~$\mu$m, as well as four polarimetric imagers covering four broad bands between 92 and 235~$\mu$m.

\begin{figure}[ht!]
\begin{center}
\includegraphics[width=\linewidth]{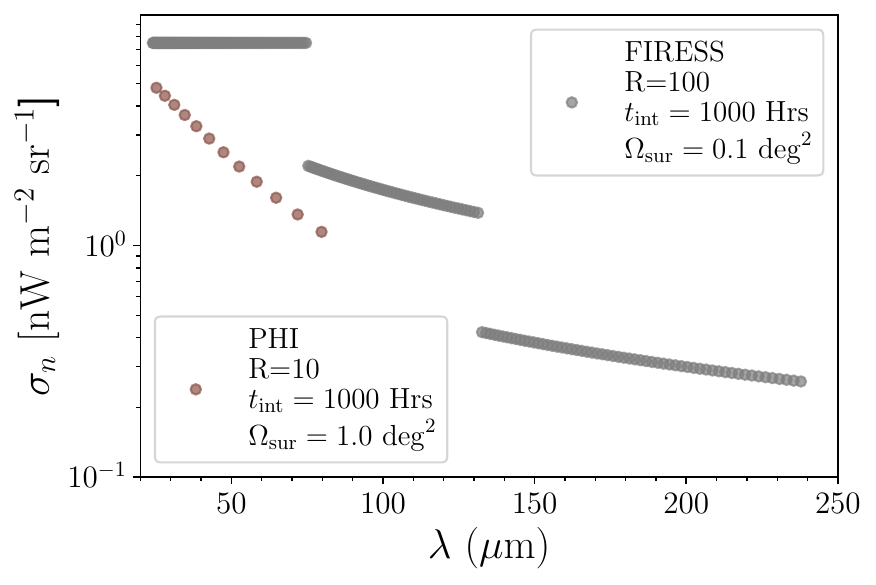}
\caption{\label{F:prima_sign}PRIMA surface brightness sensitivity per spectral channel in a wavelength-dependent beam ($5''.1/7''.6$ to $10''.4/22''.9$ from short to long wavelengths for PHI/FIRESS) with spectral resolution $R=10/100$ for PHI/FIRESS. The brown and gray points represent the PHI and FIRESS sensitivity, respectively, assuming a fiducial survey of 1000~hr of integration covering an area of $1/0.1$ deg$^2$ for PHI/FIRESS.}
\end{center}
\end{figure}

In this work, we focus on the spectral-intensity maps observed by the two PRIMAger Hyperspectral Imagers (which we refer to as PHI hereafter). The PHI beam size ranges from $5''.1$ to $10''.4$ from short to long wavelengths. With a spectral resolving power of $R=10$, there are 12 channels in total across the two hyperspectral imagers. We consider a fiducial PHI survey covering $\Omega_{\rm sur} = 1$~deg$^2$ with a total integration time of 1000~hr. The instrument noise levels are derived from the official PRIMA website\footnote{\url{https://prima.ipac.caltech.edu/page/instruments}} and shown in Figure~\ref{F:prima_sign}.

We assume a Lorentzian spectral response function with FWHM given by $\lambda/R$, where $R=10$. Since the PHI employs linear variable filters, it is possible to define finer spectral channels while maintaining the same response profile. This allows for improved spectral sampling. In this work, we consider a spectral binning of $R=20$, resulting in 24 spectral channels. The noise rms for the $R=20$ binning is scaled by a factor of $\sqrt{2}$ accordingly.

FIRESS provides two spectroscopy modes spanning 24 to 235~$\mu$m. Its low-resolution mode uses four slit-fed grating spectrometer modules with spectral resolution $R\sim100$, while a Fourier Transform Module can be inserted to provide high-resolution spectra with $R=2{,}100$--$20{,}000$. In this work, we focus on the low-resolution mode, assuming a uniform spectral resolution of $R=100$, which yields a total of 229 spectral channels between 24 and 235~$\mu$m. The FIRESS beam size varies from $7''.6$ to $22''.9$ across this wavelength range. We consider a fiducial FIRESS survey covering $\Omega_{\rm sur} = 0.1$~deg$^2$ with an integration time of 1000~hr. The noise levels per spectral channel for FIRESS are also taken from the official PRIMA website and shown in Figure~\ref{F:prima_sign}.

We adopt a multipole range of $500 < \ell < 2\times10^4$ with 10 logarithmically spaced $\ell$ bins for both the PHI and FIRESS analyses. The higher $\ell$ range is selected to adapt to the small field sizes considered in these cases. We acknowledge that these higher $\ell$ modes probe scales where the signal enters the nonlinear regime, beyond the validity of our current formalism. Modeling the effects of nonlinear clustering is deferred to future work.

\section{Signal Modeling}\label{S:signal_modeling}

\begin{figure*}[ht!]
\begin{center}
\includegraphics[width=\linewidth]{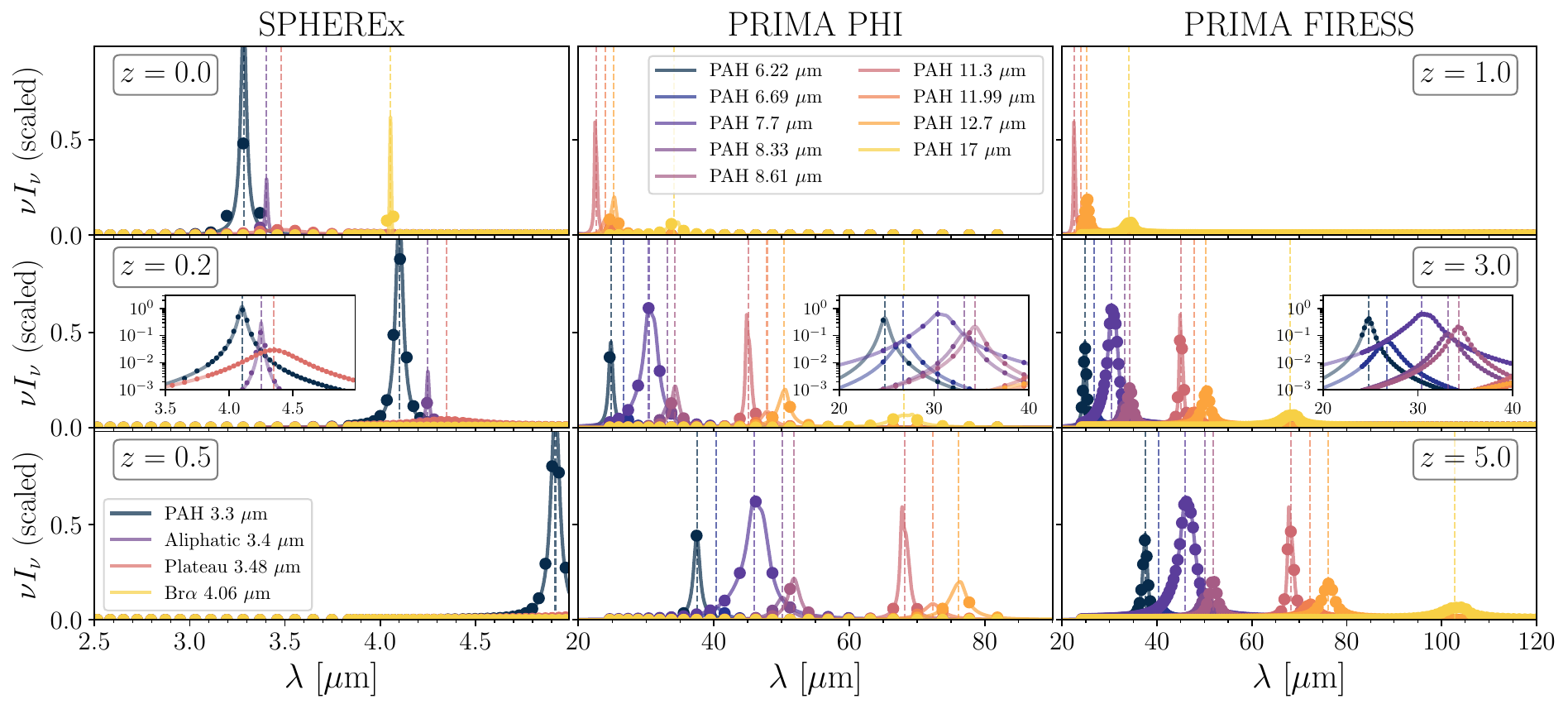}
\caption{\label{F:surveys_conv} Normalized observed spectral profiles ($\nu I_\nu$) as a function of observed wavelength for various spectral features considered in this work. The relative luminosity and spectral shape of each feature are described in Section~\ref{S:feature_models}. The left, middle, and right columns show the cases for SPHEREx, PHI, and FIRESS, respectively, with three example redshifts in each row ($z=0$, 0.2, and 0.5 for SPHEREx; $z=1$, 3, and 5 for PHI and FIRESS). Solid lines represent the intrinsic spectral shapes of the features, and dots indicate the intensities that will be observed by the spectral channels of the respective survey. The central wavelength of each feature is marked with a dashed vertical line for visual clarity. The inset panels show a zoomed-in view of the same spectrum on a log scale to better visualize the faint features.}
\end{center}
\end{figure*}

Within the observing bands of SPHEREx and PRIMA, numerous spectral lines and dust features are present, including PAH features. To apply our technique for forecasting the detectability of these features, we require a model for both their spectral profiles and luminosities. Section~\ref{S:feature_models} describes the spectral features considered in our analysis and their profiles and relative luminosities. The model for their bias-weighted luminosity density is presented in Section~\ref{S:luminosity_density_and_bias}. Finally, in Section~\ref{S:power_spectrum_model}, we use these models to compute the angular power spectra of the signals expected to be observed by SPHEREx and PRIMA.

\subsection{Feature Models}\label{S:feature_models}

\begin{deluxetable}{rllll}
\digitalasset
\tablewidth{0pt}
\tablecaption{\label{T:features}Spectral Feature Model}
\tablehead{
\colhead{Name} &\colhead{$\lambda$ [$\mu$m]} & \colhead{$L/L_{3.3}$} & \colhead{[$z_{\rm min}, z_{\rm max}$]} & \colhead{$Nz$}
}
\startdata
PAH & $3.3$ & 1.00 & [0, 0.5] & 6\\
Aliphatic & $3.4$ & 0.09 & [0, 0.4] & 5\\
Plateau & $3.48$ & 0.18 & [0, 0.4] & 5\\
Br$\alpha$ & $4.06$ & 0.08 & [0, 0.2] & 3\\
\hline
PAH & $6.22$ & 8.40 & [2.8, 5.8] & 7\\
PAH & $6.69$ & 2.72 & [2.5, 6.0] & 8\\
PAH & $7.7$ & 32.24 & [2.0, 6.0] & 9\\
PAH & $8.33$ & 3.62 & [2.0, 5.0] & 7\\
PAH & $8.61$ & 5.29 & [1.8, 5.8] & 9\\
PAH & $11.3$ & 6.51 & [1.2, 5.7] & 10\\
PAH & $11.99$ & 2.36 & [1.2, 5.2] & 9\\
PAH & $12.7$ & 4.93 & [1.2, 5.2] & 9\\
PAH & $17$ & 2.71 & [0.5, 4.0] & 8\\
\enddata
\tablecomments{Spectral features considered in this work. Features listed above/below the dividing line are those used in the SPHEREx/PRIMA (PHI and FIRESS) cases, respectively. The total luminosity of each feature is normalized to that of the PAH 3.3~$\mu$m feature. The [$z_{\rm min}, z_{\rm max}$] column denotes the range spanned by the redshift anchor points, with a spacing of $\Delta z = 0.1$ and $0.5$ for SPHEREx and PRIMA cases, respectively. The total number of redshift anchors used in each case is listed in the $N_z$ column\footnote{The PAH wavelengths listed in this table follow \citet{Smith:2007}, while the actual central wavelengths for some PAH features in our calculation is slightly offset to match the latest dataset. Specifically, the central wavelengths for PAH 3.3, 6.69, 8.33, 8.61 and 11.99 $\mu$m are 3.28, 6.70, 8.32, 8.60, 12.00 $\mu$m, respectively.}.}
\end{deluxetable}

The IM analysis employs features from the 1C starburst template presented by \citet{Lai:2020}, which is derived from combined AKARI and Spitzer observations, ranging from 5 to 38 $\mu$m. We obtained the relative luminosities of these features, including both emission lines and PAH bands, derived from the spectral decomposition of the 1C template at a resolution of $R\sim$100. Since this spectral resolution may not fully resolve these emission lines and bands, we implemented several assumptions to generate more realistic \textit{intrinsic} profiles in our model: (i) delta functions for narrow emission lines, and (ii) width corrections for broad dust features (e.g., PAH features) based on JWST's high spectral resolution spectroscopy of the starburst ring in NGC~7469 \citep{Lai2022, Lai2023}. The width correction shrinks the Drude profiles' FWHM by a factor of 2--4 for the shorter wavelength PAH features (6.22 $\mu$m and below), while the longer wavelength profiles are already well-resolved by Spitzer. Using the updated FWHM values for each PAH feature, we derived the \textit{intrinsic} PAH profiles by conserving flux under a Drude profile, following the approach of \citet{Smith:2007}. We further group blended PAH features into various complexes at 7.7, 11.3, 12.7, and 17~$\mu$m, with the individual components comprising each complex listed in Table~3 of \citet{Smith:2007}.

We do not consider potential variations in the spectral profiles from our adopted model. Since we are modeling the aggregate emission from all sources at a given redshift, rather than individual galaxies, individual variations are expected to average out statistically. Furthermore, variations in PAH spectral shapes are generally modest. Studies have shown that the central wavelengths of PAH features can vary by $\sim$0.2\% in individual Galactic sources \citep{Tokunaga:1991, Peeters:2002}, implying even smaller variations in integrated extragalactic spectra. These differences are well below the spectral resolution capabilities of SPHEREx and PRIMA.

For the SPHEREx case, in addition to PAHs, we also consider two other spectral features at 3.4~$\mu$m and 3.48~$\mu$m, as well as the nearby 4.06~$\mu$m Br$\alpha$ line. While PAH features arise from aromatic C--H stretching modes, the 3.4~$\mu$m feature originates from aliphatic C--H stretches, typically associated with carbonaceous chain structures. The broad 3.48~$\mu$m plateau feature is often attributed to aromatic carriers because it shows a slightly tighter correlation with the 3.3~$\mu$m PAH band than with the 3.4~$\mu$m aliphatic band \citep{Pilleri:2015, Lai:2020}. However, this difference is only marginal, and further studies are needed to constrain the true origin of the 3.48~$\mu$m plateau feature.

In our model, we select the four/nine brightest features for the SPHEREx and PRIMA (PHI and FIRESS) cases, respectively. These are listed in Table~\ref{T:features}, along with their relative total luminosities normalized to the 3.3~$\mu$m PAH feature. Figure~\ref{F:surveys_conv} shows the average SEDs from these features at selected redshifts. We convolve these SEDs with the instrumental filter responses of SPHEREx, PHI, and FIRESS, and also show the resulting observed signals in Figure~\ref{F:surveys_conv}.

We consider different redshift ranges for each feature in our parametrization (Section~\ref{S:parametrization}). These redshift ranges are chosen such that the emission feature falls within the observing band and yields a reasonable $S/N$ on $M_i(z)$ within that range. The selected redshift ranges for all features are listed in Table~\ref{T:features}. As described in Section~\ref{S:parametrization}, we parametrize $M_i(z)$ using a series of redshift anchor points for each feature. These anchors are equally spaced within the chosen redshift range. We adopt a redshift spacing of $\Delta z = 0.1$ for the SPHEREx case and $\Delta z = 0.5$ for the PRIMA case. The number of redshift anchors, $N_z$, for each feature is also listed in Table~\ref{T:features}. The total number of free parameters in our analysis is given by the sum of $N_z$ over all features, which gives 19 and 76 parameters for the SPHEREx and PRIMA cases, respectively.

We assume the smoothed continuum is filtered out prior to applying our method. This can be achieved either by removing the low-frequency spectral modes or by fitting and subtracting a smooth component along each line of sight. However, residuals from this filtering may remain in the data. Further, absorption features, such as the broad $\sim$3~$\mu$m water ice feature, may not be perfectly removed and stronger line features from higher redshifts, such as H$\alpha$ in SPHEREx, may fall into the observed wavelength range and contribute to the signal as well. While we ignore these complications for the purpose of demonstrating our feature intensity mapping technique, such effects should be carefully accounted for in practical applications with a proper transfer function derived from simulations.

\subsection{Luminosity Density and Bias}\label{S:luminosity_density_and_bias}
To model the bias-weighted luminosity density, we use the linear scaling relation between $L_{3.3}$ and the SFR from \citet{Lai:2020}:
\begin{equation}
\frac{L_{3.3}}{L_\odot} = 10^{6.8}~\frac{{\rm SFR}}{M_\odot~{\rm yr}^{-1}},
\end{equation}
which is derived from a sample of star-forming galaxies. The PAH intensity is expected to be dominated by star-forming population, while the contribution from quiescent galaxies will be negligible. This is because the excitation of PAH molecules requires the absorption of far-ultraviolet photons with energies between $6$ and $13.6$~eV, typically originating from regions of active star formation \citep{Leger:1984}. Observational studies with Spitzer have demonstrated that passively evolving early-type galaxies exhibiting detectable PAH features are rare, constituting $\leqslant5\%$ of the population \citep{Bregman:2006, Bressan:2006}. Therefore, their contribution to the total PAH emission is expected to be negligible.

We adopt the SFR--halo mass ($M_h$) relation from \citet{Behroozi:2013a, Behroozi:2013b}, who derived an empirical model for the average galaxy SFR as a function of $M_h$ and $z$, ${\rm SFR}(M_h, z)$, calibrated to observations. We use their publicly released\footnote{\texttt{sfh\_z0\_z8.tar.gz} file from \url{https://www.peterbehroozi.com/data.html}} model of ${\rm SFR}(M_h, z)$ over a range of $M_h$ and $z$ bins, and interpolate between these bins to construct a continuous model. With these ingredients, along with the feature luminosity ratios listed in Table~\ref{T:features}, we obtain the mean feature luminosity as a function of halo mass and redshift, $L_i(M_h, z)$. We then estimate the luminosity density and large-scale bias by integrating over the halo mass:
\begin{equation}
M_{0,i}(z) = \frac{dL_i}{dV}(z) = \int_{M_{\rm min}}^{M_{\rm max}} dM_h~\frac{dn}{dM_h}~L_i(M_h, z),
\end{equation}
\begin{equation}
b_i(z) = \frac{\int dM_h~\frac{dn}{dM_h}(M_h,z)~b_h(M_h,z)}{\int dM_h~\frac{dn}{dM_h}(M_h,z)},
\end{equation}
where we set $M_{\rm min}=10^8~M_\odot h^{-1}$ and $M_{\rm max}=10^{15}~M_\odot h^{-1}$.
$\frac{dn}{dM_h}$ is the halo mass function \citep{Watson:2013}, and $b_h$ is the halo bias \citep{Tinker:2010}. We use the implementation of the halo mass function and bias factor provided in the \texttt{colossus} package \citep{Diemer:2018}. Note that although we model $M_{0,i}(z)$ and $b_i(z)$ separately, our method that probes the clustering-scale correlations can only constrain their product, $M_i(z)$.

\begin{figure}[ht!]
\begin{center}
\includegraphics[width=\linewidth]{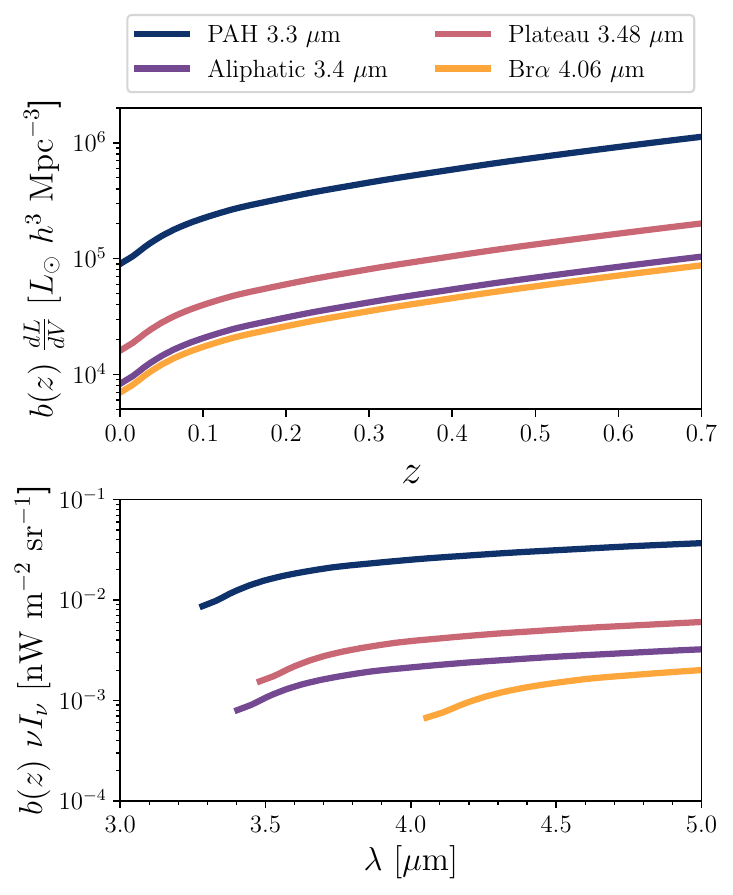}
\caption{\label{F:spherex_bdLdV}Top: Bias-weighted luminosity density as a function of redshift for the four spectral features considered in the SPHEREx case from our model (Section~\ref{S:signal_modeling}). Bottom: Bias-weighted intensity as a function of observed wavelength, derived from the bias-weighted luminosity density shown in the top panel using Equation~\eqref{E:dLdV_to_nuInu}.}
\end{center}
\end{figure}

\begin{figure}[ht!]
\begin{center}
\includegraphics[width=\linewidth]{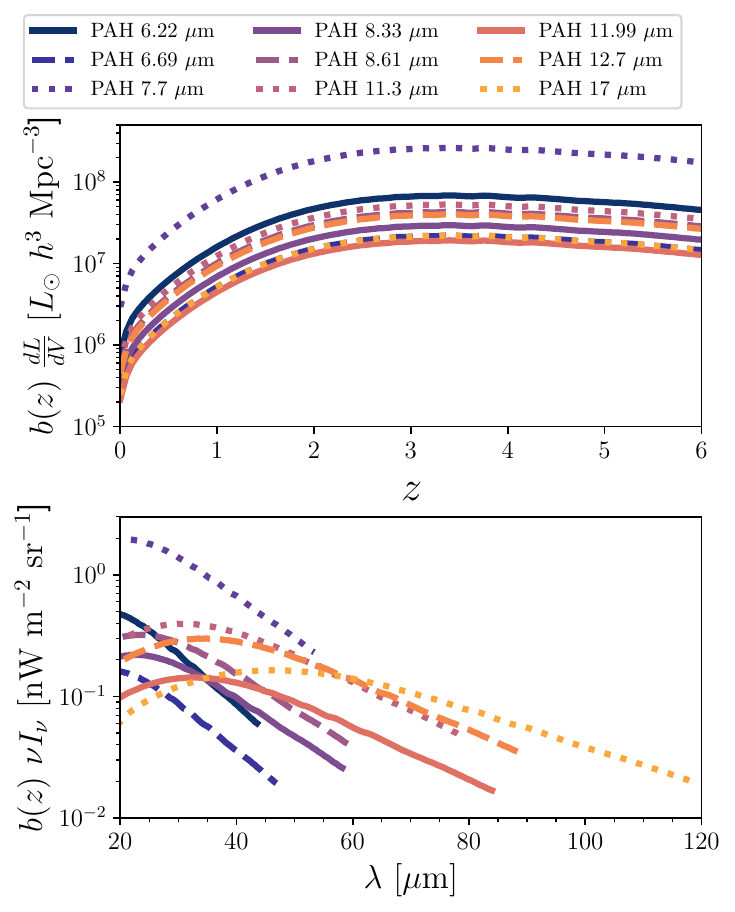}
\caption{\label{F:prima_bdLdV}Top: Bias-weighted luminosity density as a function of redshift for the four spectral features considered in the PRIMA case from our model (Section~\ref{S:signal_modeling}). Bottom: Bias-weighted intensity as a function of observed wavelength, derived from the bias-weighted luminosity density shown in the top panel using Equation~\eqref{E:dLdV_to_nuInu}.}
\end{center}
\end{figure}

Figures~\ref{F:spherex_bdLdV} and~\ref{F:prima_bdLdV} show the bias-weighted luminosity density as a function of redshift for each feature in our model for the SPHEREx and PRIMA (PHI and FIRESS) cases, respectively. In the bottom panels, we convert the luminosity density for an easier comparison to the noise levels of these surveys. For convenience, this conversion makes the simplifying assumption that the features have infinitesimal spectral width (i.e., the spectral profile is a delta function), which gives
\begin{equation}\label{E:dLdV_to_nuInu}
\nu I_\nu (\lambda_{\rm rf}(1+z)) = \left[\frac{dL}{dV}(z)\right]~\frac{c~(1+z)~D_A^2(z)}{4\pi~H(z)~D_L^2(z)}.
\end{equation}
In reality, when the observing channels are able to resolve a spectral feature across multiple channels, the observed intensity in a given channel will be the integrated contribution from a range of redshifts, weighted by the filter profile. This is effectively equivalent to applying a kernel (with the shape determined by the filter profile) to the bias-weighted intensity shown in the lower panels of Figures~\ref{F:spherex_bdLdV} and~\ref{F:prima_bdLdV}. Since this quantity varies smoothly with redshift, we expect that the actual observed bias-weighted intensity within each spectral channel will be well approximated by our simplified calculation. While the per pixel noise rms (Figure~\ref{F:spherex_R_sign} and~\ref{F:prima_sign}) is orders of magnitude higher than the expected signal in individual pixels (lower panel of Figures~\ref{F:spherex_bdLdV} and~\ref{F:prima_bdLdV}), our constraints on the aggregate extragalactic emission are derived through 3D clustering. As will be demonstrated in Section~\ref{S:results}, the large-scale correlations significantly suppress the noise, allowing for strong constraints on these emission features.

\subsection{Power Spectrum}\label{S:power_spectrum_model}
Using the spectral feature model described in Sections~\ref{S:feature_models} and~\ref{S:luminosity_density_and_bias}, along with the observing filters defined in Section~\ref{S:survey_setup}, we compute the cross-power spectrum matrices for SPHEREx, PHI, and FIRESS using Equation~\eqref{E:Cl_clus}.

\begin{figure}[ht!]
\begin{center}
\includegraphics[width=\linewidth]{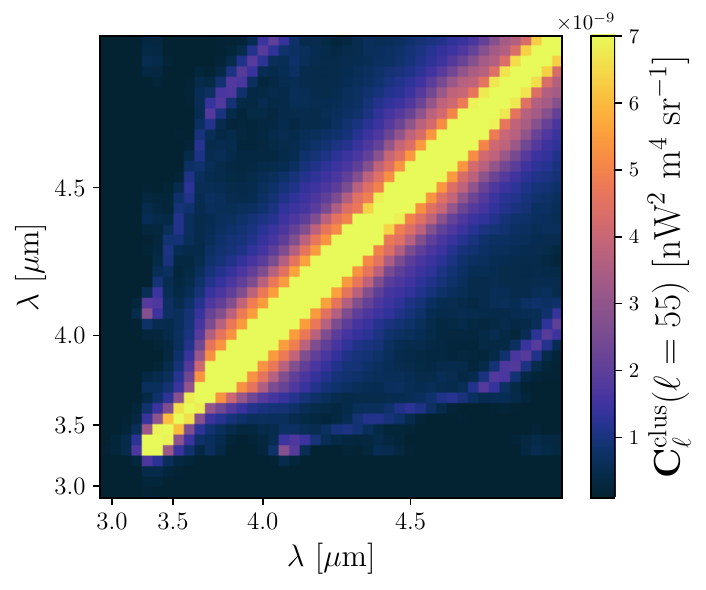}
\caption{\label{F:spherex_Cl} The clustering power spectrum matrix $C_\ell^{\rm clus}$ (Equation~\eqref{E:Cl_clus}) at the lowest $\ell$ mode, centered at $\ell=55$, for the SPHEREx case.}
\end{center}
\end{figure}

\begin{figure}[ht!]
\begin{center}
\includegraphics[width=\linewidth]{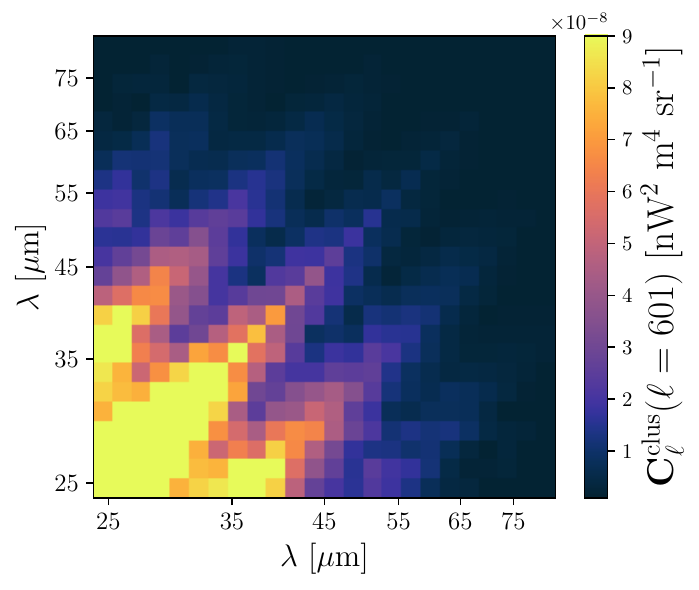}
\includegraphics[width=\linewidth]{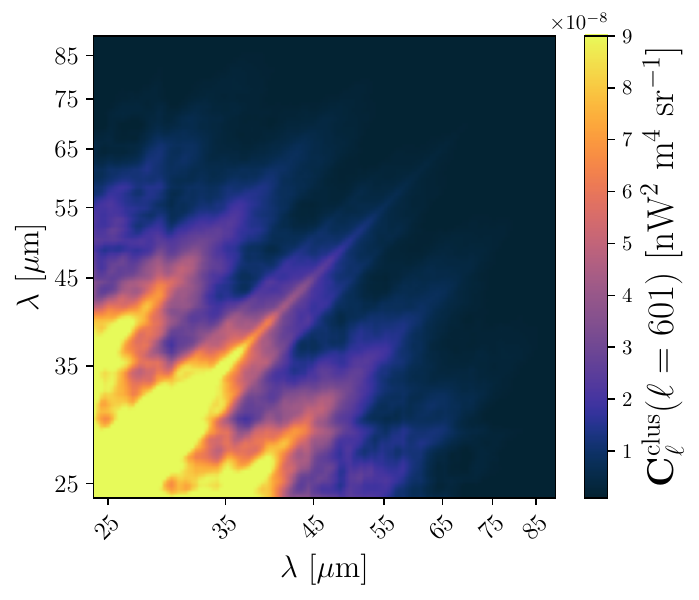}
\caption{\label{F:prima_Cl} The clustering power spectrum matrix $C_\ell^{\rm clus}$ (Equation~\eqref{E:Cl_clus}) at the lowest $\ell$ mode, centered at $\ell=601$, for the PHI (top)/FIRESS (bottom) case.}
\end{center}
\end{figure}

Figures~\ref{F:spherex_Cl} and~\ref{F:prima_Cl} show the clustering power spectrum matrix at the lowest-$\ell$ bin for SPHEREx and PRIMA, respectively. The diagonal elements represent the auto-power spectra of individual spectral channels. The band along the diagonal reflects the cross correlations between nearby channels, which are expected to be strong due to the broad spectral width of features that span multiple channels. Additional off-diagonal streaks correspond to correlations between widely separated features that trace the same redshift, leading to coherence between non-adjacent channels.

\begin{figure}[ht!]
\begin{center}
\includegraphics[width=\linewidth]{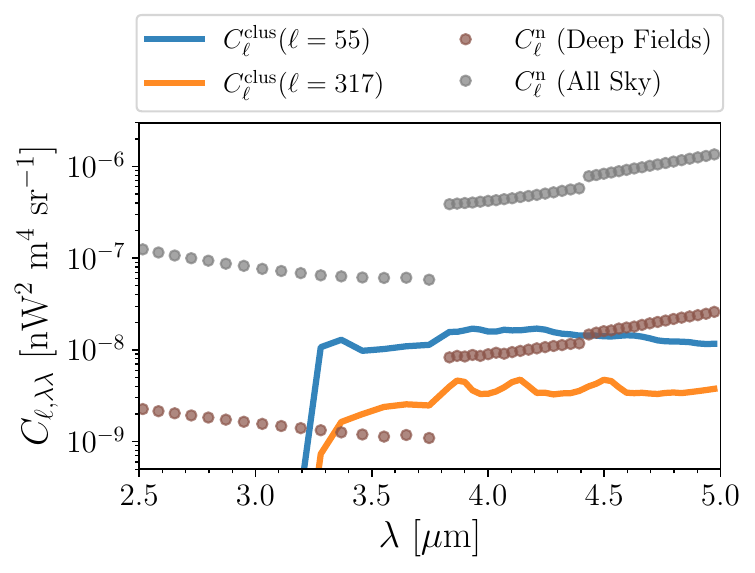}
\caption{\label{F:spherex_Cl_diag} The auto power spectrum $C_{\ell, \lambda\lambda}^{\rm clus}$---i.e., the diagonal elements of the power spectrum matrix shown in Figure~\ref{F:spherex_Cl}---for the lowest and highest $\ell$ bins centered at $\ell=55$ and $\ell=317$ in the SPHEREx case (blue and orange). The brown and gray dots show the noise power spectrum $C_\ell^{n}$ (Equation~\eqref{E:Cl_n}) for each channel in the deep-fields and the all-sky survey, respectively.}
\end{center}
\end{figure}

\begin{figure}[ht!]
\begin{center}
\includegraphics[width=\linewidth]{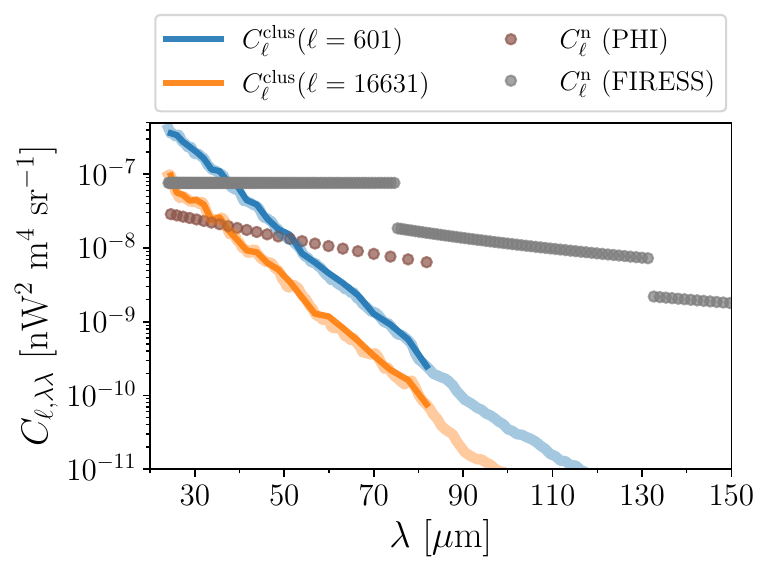}
\caption{\label{F:prima_Cl_diag} the auto power spectrum $C_{\ell, \lambda\lambda}^{\rm clus}$---i.e., the diagonal elements of the power spectrum matrix shown in Figure~\ref{F:prima_Cl}---for the lowest and highest $\ell$ bins centered at $\ell=601$ and $\ell=16631$ in the PHI/FIRESS case (blue and orange). The dark and light-colored lines show the auto power spectra for PHI and FIRESS, respectively. The brown and gray dots show the noise power spectrum $C_\ell^{n}$ (Equation~\eqref{E:Cl_n}) for each channel ($R=20$ and $R=100$) for PHI and FIRESS, respectively.}
\end{center}
\end{figure}

Figures~\ref{F:spherex_Cl_diag} and~\ref{F:prima_Cl_diag} display the auto-power spectra (diagonal elements of Figure~\ref{F:spherex_Cl} and~\ref{F:prima_Cl}) as a function of wavelength, compared to the corresponding noise power. Although the noise power dominates the auto spectra in most channels, we emphasize that noise affects only the auto spectra. In contrast, the cross-channel correlations encode critical information about the signal and are not contaminated by uncorrelated noise. Therefore, we can still extract significant information from these correlations, as demonstrated in Section~\ref{S:results}.

\section{Results}\label{S:results}
With our model of the signals---i.e., the bias-weighted luminosity density $M_i(z)$ and the spectral profile $S_i(\lambda_{\rm rf})$ for each feature $i$---from Section~\ref{S:signal_modeling}, along with the survey setup described in Section~\ref{S:survey_setup}, we compute the parameter constraints from SPHEREx and PRIMA using the Fisher formalism (Section~\ref{S:inference}). The parameters of interest are the $M_i(z)$ values at the redshift anchor points specified in Section~\ref{S:signal_modeling}. In this section, we present the parameter constraints for our fiducial signal models and survey configurations.

\subsection{SPHEREx}\label{S:results_spherex}
\begin{figure}[ht!]
\begin{center}
\includegraphics[width=\linewidth]{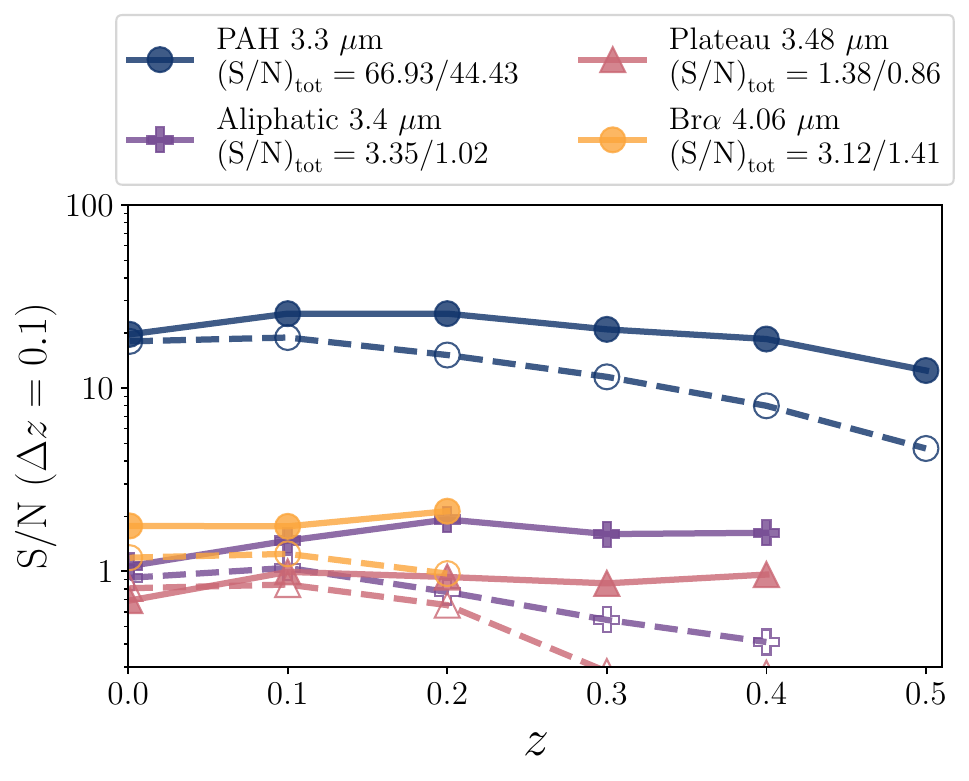}
\caption{\label{F:spherex_SNR} The $S/N$ for constraining the bias-weighted luminosity density in $\Delta z = 0.1$ bins of redshifts for each feature in the SPHEREx case. Solid/dashed lines with filled/open circles represent the deep fields/all-sky cases, respectively. The total $S/N$ of each feature summed over all redshift bins (Equation~\eqref{E:SNR_tot}) for the deep-fields/all-sky cases is provided in the legend.}
\end{center}
\end{figure}

Figure~\ref{F:spherex_SNR} displays the $S/N$ on $M_i(z)$ for the four features considered in the SPHEREx case. The total $S/N$ across all redshift bins (Equation ~\ref{E:SNR_tot}) for each feature is indicated in the legend.

We find that, despite the very different depths and survey areas between the all-sky and deep-field configurations, the $S/N$ on detecting $M_i(z)$ for the features is of a similar order of magnitude, with slightly higher sensitivity expected in the deep fields. In practice, image systematics and foregrounds are better studied in the deep fields than in the all-sky maps for SPHEREx, making the deep fields preferable for conducting this measurement compared to the all-sky survey.

Among the four features considered here, only the 3.3~$\mu$m PAH feature is expected to reach a significant detection ($>5\sigma$). The 3.4~$\mu$m aliphatic feature, the 3.48~$\mu$m plateau, and the 4.06~$\mu$m Br$\alpha$ line have total $S/N \lesssim 3$.

\subsection{PRIMA}\label{S:results_prima}

\begin{figure}[ht!]
\begin{center}
\includegraphics[width=\linewidth]{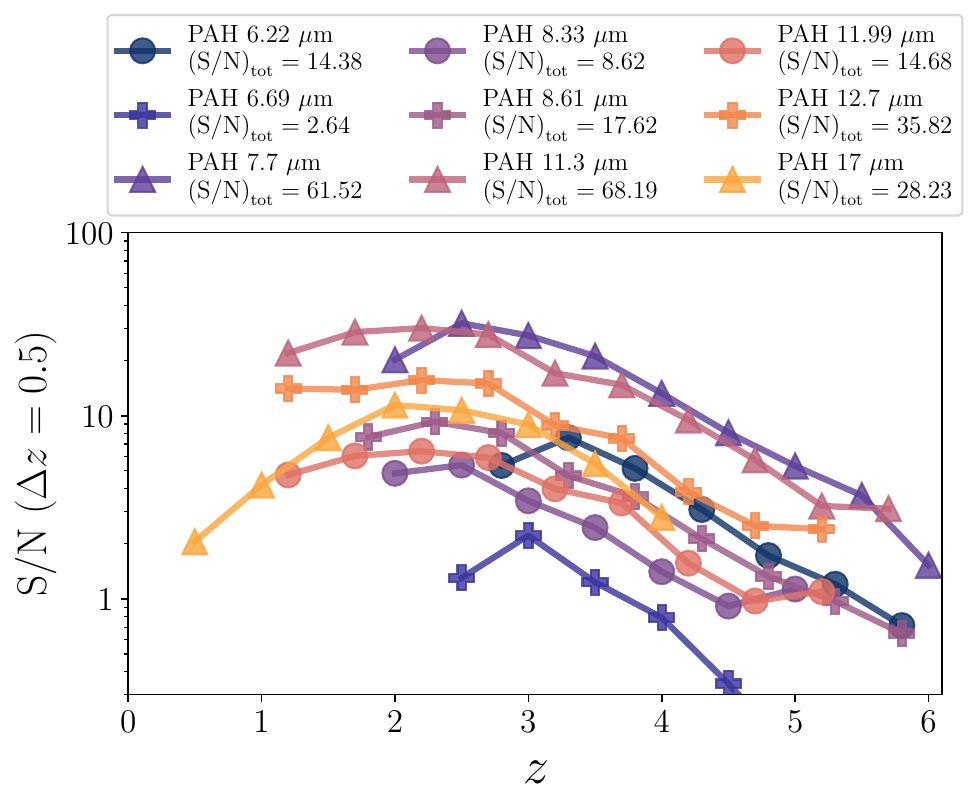}
\includegraphics[width=\linewidth]{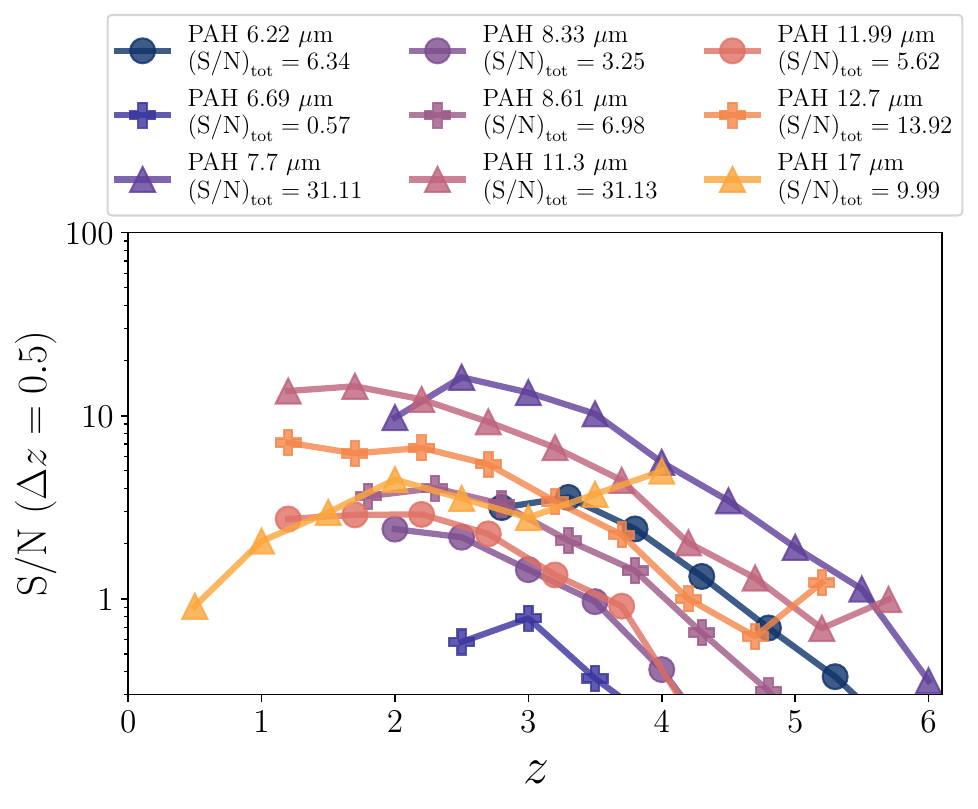}
\caption{\label{F:prima_SNR}The $S/N$ for constraining the bias-weighted luminosity density in $\Delta z = 0.5$ bins of redshifts for each feature in the fiducial PHI (top) and FIRESS (bottom) case. The total $S/N$ of each feature summed over all redshift bins (Equation~\eqref{E:SNR_tot}) is provided in the legend.}
\end{center}
\end{figure}

Figure~\ref{F:prima_SNR} displays the $S/N$ on $M_i(z)$ for the nine PAH features considered in the fiducial PHI and FIRESS cases. The total $S/N$ across all redshift bins (Equation~\ref{E:SNR_tot}) for each feature is indicated in the legend.

With 1,000-hr of integration time for both PHI and FIRESS, PHI achieves $S/N$ values approximately twice as high as FIRESS for all PAH features. The 7.7 and 11.3~$\mu$m PAHs have the highest $S/N$ because they are relatively brighter PAH features in our SED model. The next highest total sensitivities come from the 12.7 and 17~$\mu$m features. Although these are not as intrinsically bright as, for example, the 6.22~$\mu$m feature, they are observed at lower redshifts, which boosts their observed intensities. Our forecast predicts that the aggregate emission from multiple PAH features across a broad redshift range, spanning from $z \sim 1$ to $z \sim 5$, can be detected with high significance through their large-scale clustering using our feature intensity mapping technique.

\section{Discussion}\label{S:discussion}

\subsection{Intensity Mapping with Unresolved Sources}
\begin{figure}[ht!]
\begin{center}
\includegraphics[width=\linewidth]{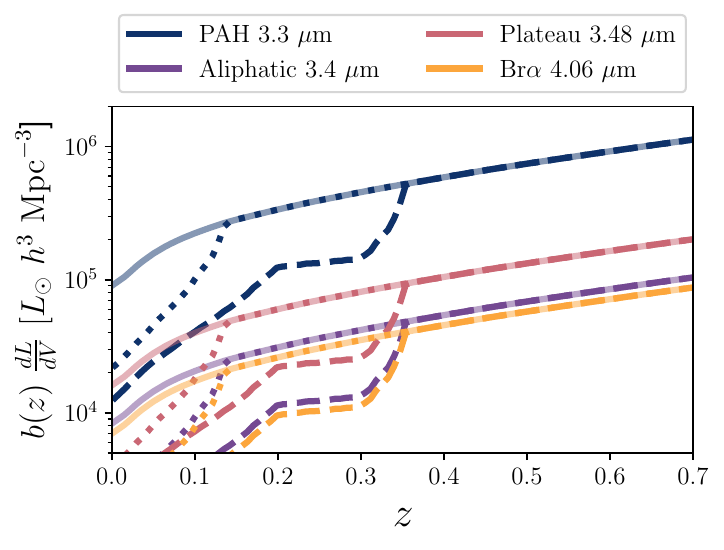}
\caption{\label{F:spherex_sfrcut_bdLdV}The bias-weighted luminosity density of the four features in the SPHEREx case. Solid lines show the signal from all sources in the fiducial case, as presented in Figure~\ref{F:spherex_bdLdV}. Dashed and dotted lines represent the cases where sources individually detectable in the SPHEREx deep-fields and all-sky survey, respectively, have been removed.}
\end{center}
\end{figure}

\begin{figure}[ht!]
\begin{center}
\includegraphics[width=\linewidth]{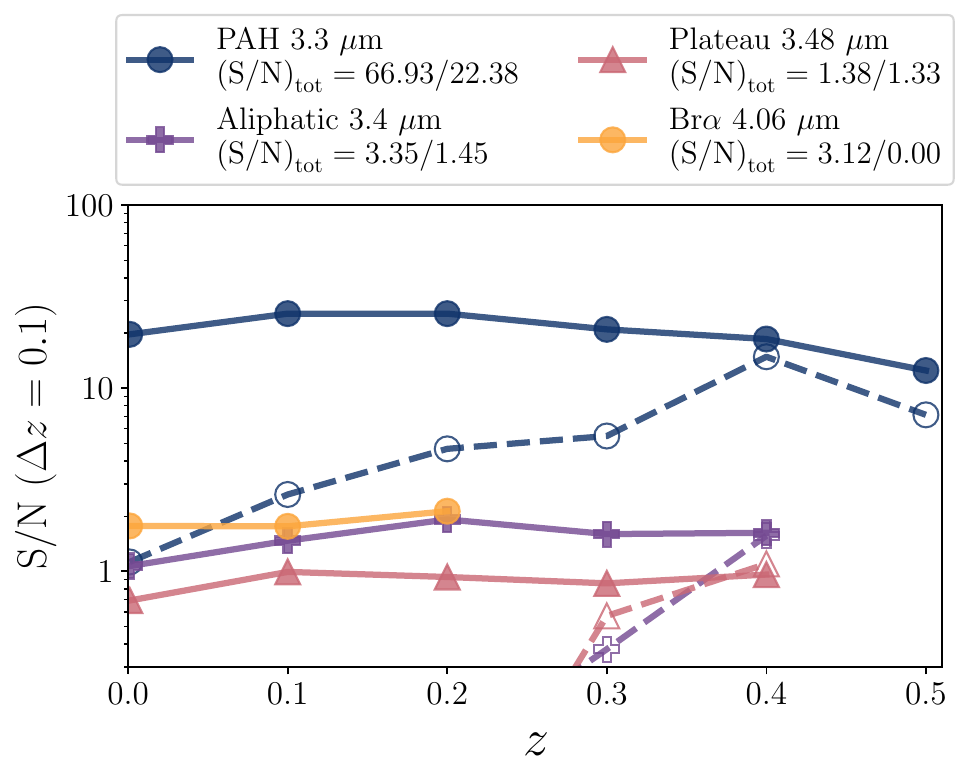}
\includegraphics[width=\linewidth]{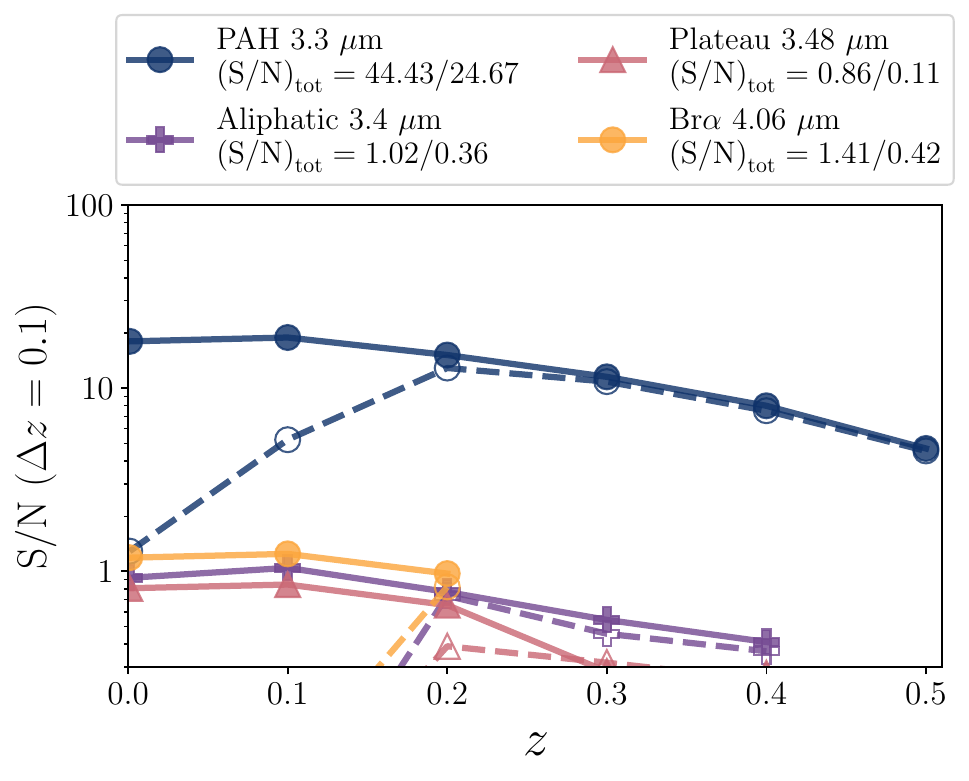}
\caption{\label{F:spherex_sfrcut} The $S/N$ for constraining the bias-weighted luminosity density in $\Delta z = 0.1$ bins for each feature in the fiducial case without masking detected sources (solid lines with filled circles), and in the case where all detected sources in the deep fields (top) / all sky (bottom) are masked. }
\end{center}
\end{figure}

In our model described in Section~\ref{S:signal_modeling}, we consider the emission from \textit{all} sources within the survey. However, brighter sources are expected to be individually detectable in these surveys. If the detection threshold is sufficiently deep such that most of the total emission is captured by resolved sources, performing IM becomes less meaningful.

A recent study by \cite{Zhang:2025} investigated the individual detectability of the 3.3~$\mu$m PAH feature in both the SPHEREx all-sky and deep-field configurations using the SPHEREx Sky Simulator \citep{Crill:2025}. They derived the SFR and $K$-band magnitude limits above which the 3.3~$\mu$m PAH feature can be individually detected. Assuming that such detected sources can be masked in the data, we estimate the bias-weighted intensity contributed only by sources below the detection threshold, and repeat our sensitivity forecasts to evaluate the detectability of the remaining unresolved emission. Specifically, we adopt the SFR limits on SPHEREx all-sky configuration from \cite{Zhang:2025}, and assume the deep-fields SFR limit to be eight times lower, corresponding to the approximated point source sensitivity improvement on deep fields compared to all sky. In this estimate, we neglect the potential signal loss and mode coupling effects introduced by masking.

The bias-weighted intensity from only the sources below the detection threshold is shown in Figure~\ref{F:spherex_sfrcut_bdLdV}. We find that, for SPHEREx, at $z \lesssim 0.15$ and $0.3$ in the all-sky and deep-field configurations, respectively, a significant fraction of the 3.3~$\mu$m PAH emission originates from galaxies that can be individually detected. Beyond these redshift limits, however, the majority of the emission comes from faint, unresolved sources below the detection limit. Figure~\ref{F:spherex_sfrcut} displays the $S/N$ on detecting the signals from only the subthreshold sources. Our method can still achieve a $\gtrsim 10\sigma$ detection of the aggregate 3.3~$\mu$m PAH intensity at these redshifts. This highlights the complementarity of our approach to individual source detection techniques.

For the case of PRIMA, there is no study analogous to \citet{Zhang:2025} that quantifies the PAH detection threshold for individual galaxies. Using a semi-analytic model and considering 1,500~hr of PHI observations, \citet{Bisigello:2024} estimated the IR luminosity and SFR distributions of detectable galaxies. Their forecast is performed on both a deep (1 deg$^2$) and a wide (10 deg$^2$) survey, each with the same 1,500~hr integration time. Our fiducial PHI case of 1 deg$^2$ with 1,000~hr integration is similar to their deep survey with a slightly shorter integration, and thus their detection limit can serve as a lower bound of the detectable IR luminosity and the SFR values. Applying their SFR limits to our model of PAH intensity, we find that at $z \gtrsim 4$, the majority ($>99\%$) of the PAH total intensity is contributed by faint, unresolved galaxies. As shown in Section~\ref{S:results_prima}, applying our feature intensity mapping technique to PHI can achieve $S/N > 3$ for multiple PAH features at this redshift range. These aggregate signals from faint sources will be uniquely accessible through IM, offering a complementary capability to individual galaxy detection with PRIMA.

\subsection{Survey Optimization}
Since PRIMA is still in the design phase and a concrete survey plan has not yet been finalized, we explore the sensitivity of PAH intensity mapping performance under different survey configurations. Specifically, assuming a fixed total integration time of 1,000~hr as our fiducial case, we vary the survey area, which effectively changes the noise level and the effective number of modes (through the $f_{\rm sky}$ factor in Equation~\eqref{E:nell}). We repeat our sensitivity forecasts for scenarios where the survey area is scaled up and down by a factor of 10 relative to our fiducial PHI and FIRESS setups. Since most of the sensitivity comes from scales much smaller than the survey size, we fix the $\ell$ range and binning when scaling the survey area, and scale the number of modes $n_\ell$ (Equation~\ref{E:nell}) proportionally to the survey area. The resulting total sensitivity for each PAH feature under these configurations is summarized in Table~\ref{T:survey_sizes}. We find that our adopted fiducial survey sizes are close to optimal for both PHI and FIRESS.

\begin{deluxetable*}{r|lll|lll}
\digitalasset
\tablewidth{0pt}
\tablecaption{\label{T:survey_sizes}Sensitivity with Different Survey Sizes}
\tablehead{
\colhead{$\lambda$} & & PHI & & &FIRESS & \\
\colhead{$\mu$m} & 0.1 deg$^2$ & 1.0 deg$^2$ & 10 deg$^2$ & 0.01 deg$^2$ & 0.1 deg$^2$ & 1 deg$^2$
}
\startdata
$6.22$ & 10.3& 14.4& 9.19& 5.06& 6.34& 4.03\\
$6.69$ & 1.41& 2.64& 1.46& 0.18& 0.57& 0.16\\
$7.7$ & 40.3& 61.5& 40.1& 21.1& 31.1& 20.9\\
$8.33$ & 6.30& 8.62& 4.64& 2.86& 3.25& 1.39\\
$8.61$ & 13.9& 17.6& 10.1& 7.03& 6.98& 4.04\\
$11.3$ & 65.8& 68.2& 39.3& 35.7& 31.1& 15.9\\
$11.99$ & 13.6& 14.7& 7.86& 5.29& 5.62& 2.84\\
$12.7$ & 35.3& 35.8& 20.4& 14.7& 13.9& 8.03\\
$17$ & 24.6& 28.2& 19.9& 11.1& 10.0& 5.89\\
\enddata
\tablecomments{The $(S/N)_{\rm tot}$ for the nine PAH features for PHI and FIRESS with 1,000~hr of integration time under different survey area assumptions. The middle case represents our fiducial survey setup, while the other two cases correspond to survey areas that are scaled by a factor of 10 smaller and larger, respectively.}
\end{deluxetable*}

\subsection{Implication of PAH Intensity Mapping}
Using the feature intensity mapping technique developed in this work, we are able to perform 3D intensity mapping for broad spectral features such as PAHs, analogous to LIM. Our forecasts demonstrate promising prospects for probing multiple PAH features across cosmic time. SPHEREx, which covers wavelengths up to 5~$\mu$m, will enable mapping of the 3.3~$\mu$m PAH feature up to $z \sim 0.5$, while PRIMA is expected to be capable of mapping multiple PAH features with high significance up to $z \sim 5$.

These measurements will offer new insights into star formation and ISM physics across cosmic history. The relative strengths of different PAH features can inform us about dust properties, such as grain size distribution and composition, as well as the interstellar environment in which the dust is embedded \citep[e.g.,][]{Maragkoudakis:2020, Draine:2021}. By accessing the galaxies beyond the most extremely luminous, PAH feature intensity mapping provides a way to quantify the buildup of PAHs in normal star-forming galaxies and to compare these timescales to the star-formation history and the metal enrichment history of the Universe over cosmic time.

Although the detectability of the 3.4 $\mu$m aliphatic feature with SPHEREx is marginal with our model, a constraint on the aliphatic-to-aromatic ratio (traced by the 3.4 and 3.3~$\mu$m intensity ratio) across a broad redshift range would enable studies of the evolution of the ISM. In particular, this ratio constrains the relative abundances of aromatic and aliphatic bonds, which are key indicators of the carbon budget and its evolution over cosmic time. The aliphatic-to-aromatic ratio is observed to decrease with increasing intensity of the far-ultraviolet radiation field \citep{Joblin:1996, Pilleri:2015}, since aliphatic bonds are more susceptible to destruction through photon-driven processing. Previous studies of the aliphatic-to-aromatic ratio outside the Milky Way have been largely limited to nearby galaxies \citep[e.g.,][]{Yamagishi:2012, Lai:2020}. With the advent of JWST, however, this diagnostic can now be extended to higher redshifts \citep[$z\sim0.2-0.5$;][]{Lyu:2025}. Our feature intensity mapping technique provides a complementary probe to the averaged measurements from all galaxies, while JWST focuses on individual targets.

In addition to auto and cross correlations within the same survey, synergy with external datasets provides powerful avenues for probing galaxy physics in the cosmic background. For example, by cross-correlating SPHEREx channel maps with galaxy catalogs, one can extract the aggregate galaxy SEDs, including continuum emission and spectral features such as lines and PAHs, as a function of wavelength and redshift \citep{Cheng:2022}. Furthermore, joint analyses of the PAH background mapped by SPHEREx or PRIMA with LIM surveys targeting different spectral lines (e.g., [\ion{C}{2}] or CO) can give a more comprehensive census of galaxy physics by simultaneously probing multiple phases of the ISM through distinct tracers \citep{Sun:2019}.

In addition, similar to LIM, PAHs can serve as tracers of the 3D density field, providing a novel probe of LSS at redshifts well beyond the reach of traditional galaxy surveys.

\subsection{Future Work}
The feature intensity mapping technique developed in this work is timely for the recently launched SPHEREx mission and the proposed PRIMA mission, both of which will conduct large-scale spectral-imaging surveys in the IR, where multiple PAH features fall within their wavelength coverage.

In this work, we explore the sensitivity of mapping PAH features with our technique in SPHEREx and PRIMA under nominal survey setups, while ignoring some realistic complications, such as foreground systematics, residual continuum fluctuations, and unaccounted spectral features. Future investigations will assess the robustness of this method against more realistic simulations.

We focus on PAH emission as a use case for our feature intensity mapping technique. However, this approach can be applied to mapping any broad emission feature across the electromagnetic spectrum. For example, the anomalous microwave emission has a broad spectral profile centered at $\sim 30$~GHz, which is thought to originate from spinning dust grains \citep{Dickinson:2018}. In addition, our method can be applied to absorption spectra, such as the dust absorption feature at 2175~\AA, as well as the near- and mid-infrared absorption spectra from biogenic molecular ices, including water (H$_2$O), carbon dioxide (CO$_2$), carbon monoxide (CO) and methanol (CH$_3$OH) \citep{Hollenbach:09,Boogert:11, Boogert:13, Oberg:11}.

This work is the third in our series on performing inference using the full covariance of spectral-imaging surveys, building upon \citet{Cheng:2023} and \citet{Cheng:2024}. In \citet{Cheng:2023}, we focused on reconstructing the principal modes of the continuum from cross-correlations in broadband photometric intensity maps. In \citet{Cheng:2024}, we explored the application of the same method for line intensity mapping. In this work, we extend our model to include broad spectral features such as PAHs. 

A natural next step will be to combine the methods developed thus far into a unified framework that models the full aggregate source SED, including continuum, lines, and broad spectral features, and jointly constrains all these components and their redshift evolution from spectral intensity maps. We leave this to future work.

\section{Conclusion}\label{S:conclusion}
In this work, we develop ``feature intensity mapping,'' a technique that generalizes the LIM formalism to probe the 3D clustering of the intensity field arising from broad spectral features rather than narrow emission lines. By modeling the full set of auto- and cross-power spectra between all pairs of spectral channels in a spectral-imaging survey, and forward-modeling the convolution of the intrinsic spectral profiles with the instrumental bandpasses, we present a Bayesian inference framework to simultaneously recover the bias-weighted intensity of multiple spectral features as a function of redshift.

We introduce the formalism of feature intensity mapping and apply it to forecast the detectability of PAH emission in the intensity mapping regime with SPHEREx and proposed PRIMA observations. We model several PAH features as a function of redshift based on scaling relations with SFR derived from previous studies, and adopt their spectral profiles from existing spectroscopic measurements. While SPHEREx’s spectral coverage includes only the 3.3\,$\mu$m PAH feature at $z\lesssim0.5$, PRIMA will be capable of probing multiple PAH features across a much broader redshift range, up to $z\sim5$.

For the SPHEREx forecast, we include the 3.3\,$\mu$m PAH feature along with two nearby fainter features at 3.4 and 3.48\,$\mu$m, as well as the Br$\alpha$ line, and apply our method to both the all-sky and deep-field configurations. We find that the 3.3\,$\mu$m PAH feature can be detected at high significance ($S/N \gtrsim 10$ per $\Delta z = 0.1$ bins of redshifts) in both survey modes over the nominal two-year mission, with the deep fields achieving slightly higher sensitivity due to lower instrumental noise, despite the smaller survey area.

For PRIMA, we consider nine PAH features spanning $6.2$ to $17\,\mu$m and redshifts $0.5 \lesssim z \lesssim 5$. We assume PHI and FIRESS surveys covering 1 and 0.1 deg$^2$, respectively, each with 1,000~hr of integration. Our forecasts show that multiple PAH features can achieve $>10\sigma$ detections of their bias-weighted intensity per $\Delta z = 0.5$ bins of redshifts at $z > 1$, with the two bright PAH features (7.7 and 11.3\,$\mu$m) detectable out to $z \sim 5$.

While individually detected galaxies at high redshift will be biased toward the brightest sources, our results demonstrate that IM offers a powerful complementary approach to recover the integrated emission from the full galaxy population, including faint and unresolved sources. Our feature intensity mapping technique enables the extension of 3D IM, previously limited to narrow emission lines with LIM, to general broad spectral features such as PAHs. This is especially crucial given the importance of PAHs in tracing star formation, and dust and metallicity evolution in galaxies. The development of this technique is timely: SPHEREx has just begun its survey, and, if selected for implementation, PRIMA could launch early in the next decade.

\begin{acknowledgements}
We thank Matt Bradford, Andreas Faisst, J. D. Smith, the SPHEREx Science Team, and the participants in ``PRIMA and the Future of Far-Infrared Science Working Meeting'' for helpful conversations. Part of this work was carried out at the Jet Propulsion Laboratory, California Institute of Technology, under a contract with the National Aeronautics and Space Administration (80NM0018D0004).
\end{acknowledgements}

\software{
astropy \citep{AstropyCollaboration:2013,AstropyCollaboration:2018,AstropyCollaboration:2022}, COLOSSUS \citep{Diemer:2018}, JAX \citep{jax2018github}
}

\vspace{5pt}
\bibliography{PAH, Planck_bib}{}
\bibliographystyle{aasjournalv7}

\end{CJK*}
\end{document}